\documentclass[12pt,preprint]{aastex} 
\usepackage{graphicx}
\usepackage{array}
\usepackage{amsmath}
\usepackage{amsbsy}

\renewcommand{\vec}[1]{\ensuremath{\boldsymbol{#1}}}
\newcommand{\mat}[1]{\ensuremath{\boldsymbol{#1}}}

\begin{document}

\title{Time-Dependent Reconstruction of Non-Stationary Objects with Tomographic
or Interferometric Measurements}

\author{R.A. Frazin, M.D. Butala}
\affil{Dept. of Electrical and Computer Engineering, University of Illinois, Urbana, IL 61801}
\author{A. Kemball}
\affil{National Center for Supercomputing Applications, University of Illinois, Urbana, IL 61801}
\author{F. Kamalabadi}
\affil{Dept. of Electrical and Computer Engineering, University of Illinois, Urbana, IL 61801}

\begin{abstract}
 
In a number of astrophysical applications one tries to determine the
two-dimensional or three-dimensional structure of an object from a
time series of measurements.  While most methods used for
reconstruction assume that object is static, the data are often
acquired over a time interval during which the object may change
significantly.  This problem may be addressed with time-dependent
reconstruction methods such as Kalman filtering which models the
temporal evolution of the unknown object as a random walk that may or
may not have a deterministic component.  Time-dependent
reconstructions of a hydrodynamical simulation from its line-integral
projections are presented.  In these examples standard reconstructions
based on the static assumption are poor while the Kalman based
reconstructions are of good quality.  Implications for various
astrophysical applications, including tomography of the solar corona
and radio aperture synthesis, are discussed.

\end{abstract}

\section{Introduction}

Tomographic reconstruction has a number of astrophysical applications
such as determining the three-dimensional structure of the solar
corona and heliosphere with radio scintillation, white-light or EUV data 
(Dunn et al. 2005, Frazin 2000, Frazin \& Janzen 2002, Frazin et al.\ 2005, Hayashi et al. 2003), Doppler tomography of accreting flows (e.g., Marsh 2005), air-column tomography in adaptive optics (e.g., Tyler
1994, Tokovinin et al.\ 2001), and time-distance helioseismic
tomography of near-surface features (e.g., Duval \& Gizon 2000).  It
is quite often the case that the object exhibits significant evolution
during the time in which the data required for the inversion are collected.
For example, white-light tomography of the solar corona requires
$180^\circ$\ of rotation, which takes about two weeks.  During some
two week time periods the large scale topology of the structure may
not change a lot, but during others it can change dramatically, which
affects the quality of the reconstructions (Butala et al.\ 2005).
Using two or more spacecraft has been shown to improve the situation,
but the problem demands tomographic inversion methods which explicitly
account for temporal variations (Frazin \& Kamalabadi 2005).

Determining the time-dependent structure of an object with tomographic
methods is an example of what is sometimes called dynamic estimation
in the signal processing and statistical literature.  Dynamic
estimation also has important applications in radio astronomy
imaging. Standard approaches to radio-interferometric image formation
(Thompson et al. 2001) typically assume the source brightness
distribution under reconstruction has negligible time-variability over
the course of the observing period.  There are astrophysical sources
for which this standard assumption in Fourier synthesis imaging cannot
be made.  Studies of highly time-variable solar phenomena (Bastian
1989), comets colliding with planets (de Pater \& Brecht 2001), and
relativistic jet sources (``microquasars") in the Galaxy (Vermeulen
1993; Mioduszewski 2001) may show proper motions or flux density
variability within the span of a given observing run.  Dynamic
estimation offers the possibility of improved image fidelity in
radio-interferometric imaging of objects of this type.  Historically,
several approaches have been taken to mitigate this problem, including
sub-dividing the measured visibility data into snapshot time
intervals, over which the standard assumption of a constant source
brightness distribution has greater validity. These data segments can
then be imaged separately, but at the cost of reduced image fidelity
as a result of sparser u-v plane coverage and lower sensitivity in
each individual interval.  Furthermore, this approach does not take
advantage of the fact that an image at one measurement time is closely
related to the image at the next measurement time.

Dynamic estimation is closely related to data assimilation.  Data
assimilation incorporates physics-based models for the temporal
evolution and is commonplace in the atmospheric science, weather
prediction, oceanographic and other communities (e.g., Ghil 1989,
Sepp\"anen et al.\ 2001, Bertino et al.\ 2002, Buehner \&
Malanotte-Rizzoli 2003). The problem addressed here is that of dynamic
estimation when no physics-based model for the temporal evolution is
available.

On the surface, it may seem that the task of determining a
three-dimensional object (2 spatial dimensions plus time) from two
dimensions of data (1 spatial dimension plus rotation angle) in the
case of interferometry, or a four-dimensional object from three
dimensions of data in the case of tomography, would be impossible to
do in a useful way.  Even the static reconstruction problem is often
somewhat underdetermined or ill-conditioned, let alone the
time-dependent problem.  It is helpful to recall that the reason the
static reconstruction problems are solvable is because real objects
are smooth or can be approximated reasonably with smooth objects
(possibly allowing for edges and ridges).  Similarly, a dynamic object
tends to change in a smooth way.  Thus, many real dynamic objects can
be usefully approximated by smooth objects that vary smoothly in time.
Smoothly evolving smooth objects exist in a much smaller subspace than
is available to arbitrary objects that need not fulfill any
constraints.  The Kalman filter approach presented below effectively
finds the most smoothly evolving smooth object that matches the data.

Dynamic tomography without physics-based models has received some
attention in the literature and various methods have been proposed and
demonstrated.  Some methods are based on Kalman-type filters (e.g.,
Vauhkonen et al.\ 2001, Kolehmainen et al.\ 2003) while others are not
(Schmitt \& Louis 2002, Schmitt et al.\ 2002, Zhang et al.\ 2005).
There is also a large literature on handling patient movement and
quasi-periodic cardiac motion in medical tomography that can be found
in the IEEE Transactions on Medical Imaging.  The results given this
letter are based on the Kalman filter formulation described in the
next section.

\section{Time-Dependent Image Reconstruction with the Kalman Filter}

Kalman filtering has been used for time-dependent estimation problems
(such as tracking satellites) since its introduction in 1960 (e.g.,
Kalman 1960; Anderson \& Moore 1979; Tapley et al.\ 2004).  The Kalman
filter is based on the so-called state-space formulation, which
consists of a time-update equation:
\begin{equation}
\vec{x}_{t+1} = \mat{U}_t \vec{x}_t + \vec{g}_t,
\label{time-update}
\end{equation}
and a measurement equation:
\begin{equation}
\vec{y}_t = \mat{A}_t \vec{x}_t + \vec{n}_t,
\label{measurement}
\end{equation}
where $t$\ is a time index, and the vector $\vec{x}_t$ is a discrete
representation of the unknown object (e.g., the electron density of
the solar corona) at time $t$.  The matrix $\mat{U}_t$ is called the
update operator, $\vec{A}_t$ is the measurement operator for all
measurements made at (or suitably close to) time $t$, and $\vec{y}_t$
is the vector of measurements taken at time $t$ (e.g., a polarized
brightness image from a coronagraph).  The vector $\vec{x}_t$ is also
called the state of the system at time $t$.  The Gaussian zero-mean
vector $\vec{n}_t$ represents noise in the measurements, and the
Gaussian zero-mean vector $\vec{g}_t$, called the state noise,
represents a noise process that drives non-deterministic changes in
the state from one time to the next.  It is the state noise that gives
the model its random walk character and its smooth time evolution.  In
data assimilation the update operator $\mat{U}_t$ contains knowledge
of the physics of the system (e.g., a hydrodynamical model) and is
responsible for the deterministic character of the system behavior.
When no good physics model is available $\mat{U}_t$ is usually just
the identity operator, but it can also be used to perform simple
deterministic tasks, such as applying a differential rotation curve in
the case of solar tomography.  The Kalman formulation also requires
knowledge of the covariance matrices $\overline{\vec{g}_k
\vec{g}^T_l}$, $\overline{\vec{n}_k \vec{n}^T_l}$, and
$\overline{\vec{g}_k \vec{n}^T_l}$, where $k$\ and $l$\ are time indices (the $^T$
superscript is the transpose operator and the overline represents
averaging over the statistical ensemble).  Kalman filtering allows one
to make an estimate, $\hat{\vec{x}}_{t|t'}$, of the state at any time
step $t$\ given data up to and including time step $t'$, where $t'$
may be greater than (``smoothing''), less than (``predicting''), or
equal to (``filtering'') $t$.

When solving underdetermined or poorly conditioned linear systems of
equations it is common practise to regularize the problem (Karl 2000,
Frazin \& Janzen 2002) to avoid spurious, high-frequency oscillations
in the solution.  Regularization creates smooth solutions.  Most
treatments of the Kalman filter do not incorporate regularization, but
our simulations were much improved with regularization.  Following
Baroudi et al.\ (1998), we regularized the solution by augmenting the
measurement equation.  Instead of measurement equation
(\ref{measurement}), we used the following augmented system:
\begin{equation}
\left( \begin{array}{c} \vec{y}_t \\ \boldsymbol{0} \end{array} \right) = 
\left[ \begin{array}{c} \mat{A}_t \\ \vec{R} \end{array} \right] \vec{x}_t +  
\left( \begin{array}{c} \vec{n}_t \\ \vec{h} \end{array} \right)\!,
\label{measurement2}
\end{equation}
where $\boldsymbol{0}$ is a vector of zeros with the appropriate
length and $\vec{R}$ is a smoothing matrix, which was taken to a
finite different approximation of the gradient operator.  The vector
$\vec{h}$ is a noise process.  It is assumed that $\overline{\vec{n}_t
\vec{h}^T} = 0$, and the covariance matrices of $\vec{n}_t$ and
$\vec{h}$ must be specified.  They will be denoted as
$\mat{C}_{\vec{n}} =
\overline{\vec{n}_t \vec{n}_t^T}$ and $\mat{C}_{\vec{h}} =
\overline{\vec{h} \vec{h}^T}$, respectively.  Rigorously, the positive
definite matrix $\mat{C}_{\vec{n}}$ should include pixel-to-pixel
correlations in observational errors, but the most important aspect is
that the diagonal elements are set to the proper noise level.  The
matrix $\mat{C}_{\vec{h}}$ may take any positive definite form.  For
the examples below we used $\mat{C}_{\vec{h}} = \sigma^2
\boldsymbol{1}$, where $\boldsymbol{1}$ is the identity matrix, and
$\sigma$ is a regularization parameter.  To understand the effect of
this regularization parameter, consider the weighted least-squares
solution of equation (\ref{measurement2}) under the assumption
Gaussian noise.  The solution is given by the minimum of the following
function: $\Phi(\vec{x}_t) = (\vec{y}_t - \mat{A}_t \vec{x}_t)^T
\mat{C}_{\vec{n}}^{-1}(\vec{y}_t - \mat{A}_t
\vec{x}_t) + (1/\sigma^2) \vec{x}_t^T \mat{R}^T \vec{R} \vec{x}_t$
(Moon \& Sterling 2000, Frazin \& Kamalabadi 2005).  This is the
standard form for a static, regularized least-squares solution given
only the data $\vec{y}_t$.  When $\sigma$ is small the solution will be
highly smoothed and the data misfit vector $(\vec{y}_t - \mat{A}_t
\vec{x}_t)$ will be large.  Thus, $\sigma$ serves as a
regularization parameter which may be chosen via methods such as
cross-validation (Karl 2000, Frazin \& Janzen 2002, and references
therein).

In addition, we were able to improve the solution by adopting a
non-diagonal form of the state noise covariance matrix
$\mat{C}_{\vec{g}} = \overline{\vec{g}_t \vec{g}_t^T}$, which must be
done with care so as to ensure that it is positive definite.  In the
simulations presented below, we set $\mat{C}_{\vec{g}}$\ so that the
correlation in the state noise between any two pixels decreased
exponentially with the square of the distance between them.  The size
of the exponential decrease was characterized by a correlation length
$l$.

The formulae for the estimates $\hat{\vec{x}}_{t|t'}$\ can be found in
standard texts such as Anderson \& Moore (1979), Kailath et al.\
(2000) and Tapley et al. (2004).  For numerical stability, all
results given below were calculated with the so-called ``square-root"
or ``array" form of the Kalman smoother.  All reconstructions given in
the next sections are estimates based on all of the data, i.e., the
dynamic reconstructions shown below are elements of the sequence
$\hat{\vec{x}}_{t|N}$, where $t$\ is the time index of the image and
$N$\ is the total number of observation times.  These are the
so-called ``smoothed" estimates in the literature because $N > t$
(except for the last estimate $\hat{\vec{x}}_{N|N}$), which has little
to with smoothing in the regularization sense.

\section{Simulation Results}

In order to demonstrate the principles given above we made tomographic
reconstructions of a highly time-variable two-dimensional (2D) object
from its line-of-sight projections.  Note that this problem considered
here is equivalent to aperture synthesis of a circumpolar source with
a linear array (Bracewell \& Riddle 1967).  The time-variable object
was given by a 2D MHD simulation of a magnetized molecular cloud
collapse by C.F. Gammie and C. Ditsworth.\footnote{The 2D cloud
collapse simulation can be found on the World Wide Web at:
\url{http://ddr.astro.uiuc.edu/ddr/twod}.}  This is a highly dynamic
simulation with $N = 128$ time-steps or movie frames.  At the beginning of the
simulation the cloud is uniform with sinusoidal density and magnetic
field perturbations superimposed.  As the simulation progresses, the
cloud collapses into several strands which begin to orbit around each other as
they each undergo further collapse to form protostars by the final
time-step.

The simulated data were given by line-of-sight integrals of the MHD
movie frames.  Between each MHD time-step, the view-angle was
increased by $5.6^\circ$, so that every 64 frames the view angle
cycles through $360^\circ$, making 2 complete rotations during the
simulation period.  Since data taken at an angle $\theta$ and $\theta
+ 180^\circ$\ are redundant, 32 projections (one per time-step) were
needed to make a static reconstruction with full angular coverage.
The first half of the movie shows the object undergoing relatively
slow changes, while the second half contains most of the rapid
dynamics.  White, Gaussian noise was added to the projections such
that $\vec{y}$ had a signal-to-noise ratio (SNR) of $5\times 10^3$,
where $\mbox{SNR} = \overline{\vec{y}} / \sigma_n$.

\begin{figure}
\begin{center}
\includegraphics[width=\textwidth]{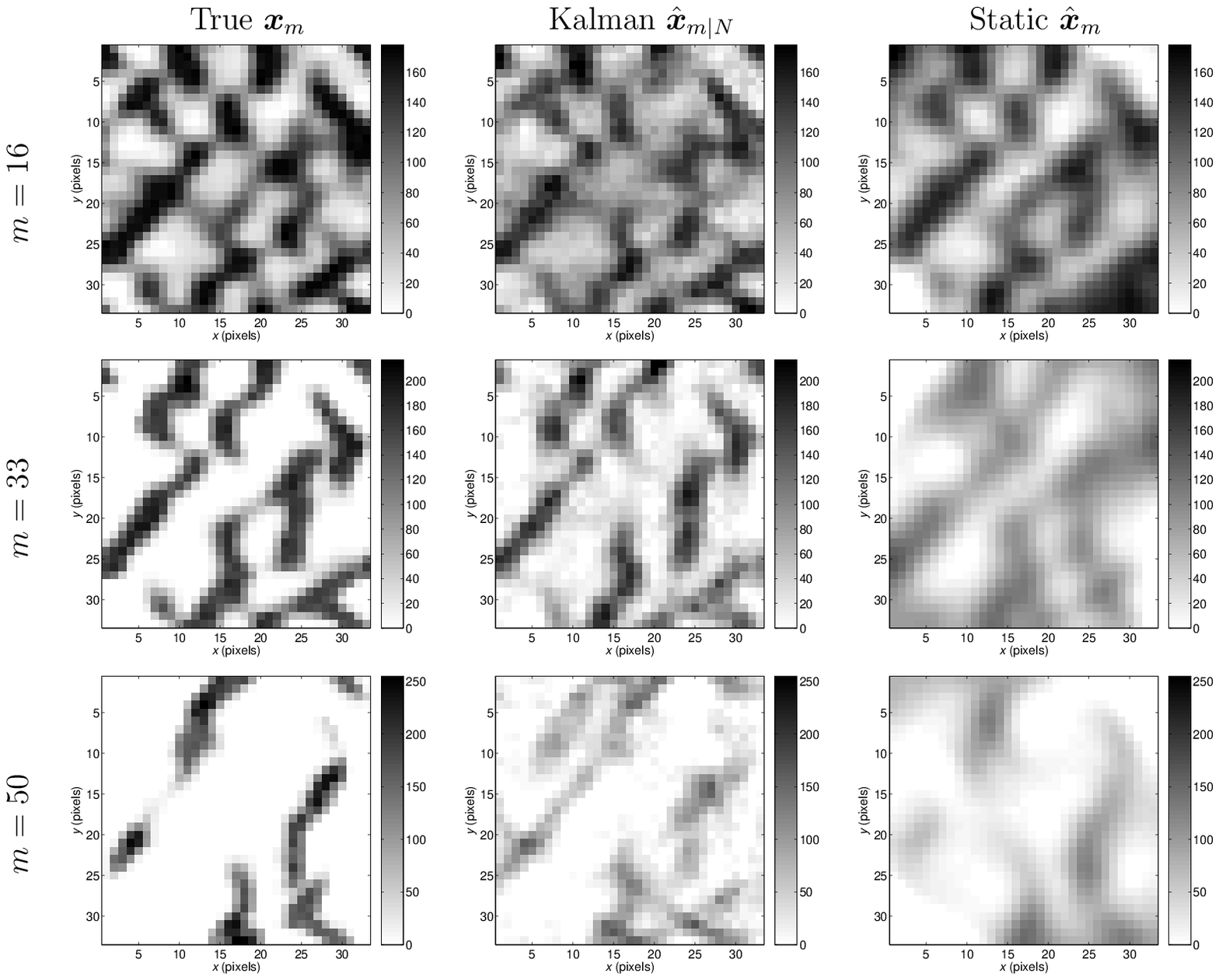}
\end{center}
\caption{\label{fig:results} A comparison between the static and
dynamic reconstructions of several frames of a 2D MHD simulation of a
self-gravitating, magnetized cloud.  The above panel is divided into
three rows, each associated with a time index $m$ of the 2D MHD
simulation.  All images in a given row are displayed using the same
color scale as the 2D MHD simulation frame of that row.  The columns
of the above panel, from left to right, contain: frames from the 2D
MHD simulation, Kalman reconstructions, and static reconstructions.
Note the superiority of the Kalman over the static reconstructions (in
particular, the static reconstructions are too smooth).  In both
cases, the quality of the reconstructions decreases as the
time-variability in the frames of the 2D MHD simulation increases.}
\end{figure}

Figure \ref{fig:results} shows the results of the simulations.  Each
row represents a different time in the MHD movie.  The first column
shows the density of the MHD movie at the time step $m$ (indicated in
the figure), the second column shows the Kalman reconstruction
$\hat{\vec{x}}_{m|N}$, and the third column shows the static
reconstruction based on 32 frames made with data centered on that time
step.  The state noise correlation length $l$, state noise amplitude
$\sqrt{L(\mat{C}_g)}$, and regularization parameter $\sigma$ where all
chosen ``by eye" to give the best looking reconstructions.  The
``optimal'' parameters were: $l$=3.6 pixels, $\sqrt{L(\mat{C}_g)}$ =
$5\times 10^{-3}\, \overline{\vec{x}}$, and  $\sigma = 0.02 \,
\sigma_n \,
L(\mat{R})/L(\mat{A_0})$, where $L(\cdot)$\ indicates that the
largest singular value of the matrix argument is to be taken.  More
rigorous methods such as cross-validation may also be used to
determine reconstruction parameters (Karl 2000, Frazin \& Janzen 2002,
and references therein).  For the static reconstructions, the
regularization parameter was chosen the minimize the difference
between the state estimate and the true state, averaged over the 32
frames used to form each static reconstructions.

\begin{figure}
\begin{center}
\includegraphics[width=0.4\textwidth]{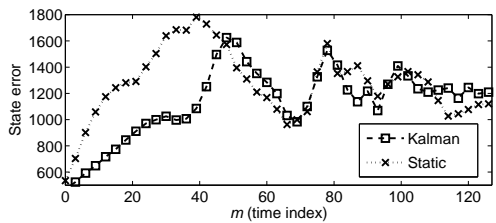}
\end{center}
\caption{\label{fig:error} A comparison between the Kalman and static
reconstruction.  The Kalman reconstructions are superior to
the static ones, except above time $m=55$\ where the object begins to change very quickly (in which case the two methods have equal errors).}
\end{figure}

The results presented in Figure~\ref{fig:results} show that the Kalman
reconstructions are much more faithful to the original than the static
reconstructions.  More quantitatively, Figure~\ref{fig:error} provides
a comparison between the Kalman and static reconstruction error,
$e_m$.  In the Kalman case, $e_m$ is defined by: $e_m =
\left\|\vec{x}_m - \hat{\vec{x}}_{m|N} \right\|$, with an analogous
definition for the static error.  At times greater than
$m = 66$\ the MHD simulation begins to exhibit extremely rapid
temporal changes, and both the Kalman and static reconstructions match
the simulation frames poorly.  However, the results indicate that in
the regime of low to moderate temporal variability, the Kalman
reconstructions are far superior.  Quantifying the effects of temporal
variability is a subject of ongoing research.  Note that the far
superior qualitative morphology of the Kalman over the static
reconstructions evident in Figure~\ref{fig:results} does not necessary
result in a far superior error in Figure~\ref{fig:error}.  The
quantification of reconstruction quality is also a topic of continuing
study.

\begin{figure}
\begin{center}
\includegraphics[width=0.8\textwidth]{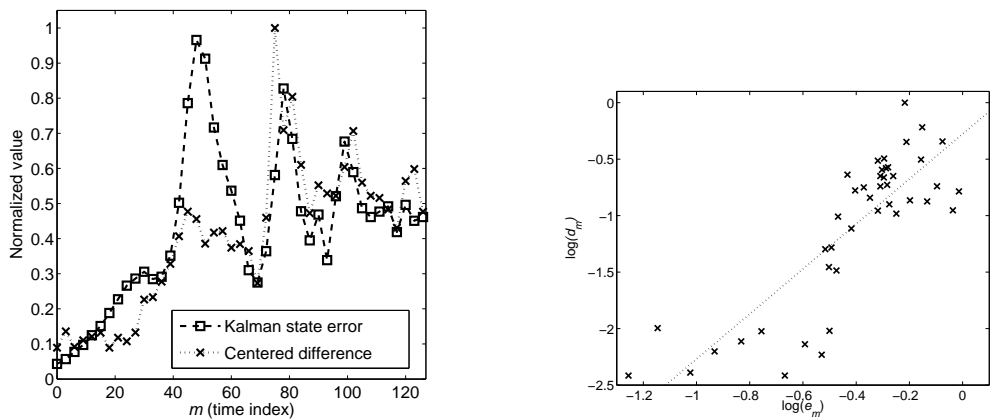}
\caption{\label{fig:e_d_comp}Left: The normalized magnitude of
the dynamic estimation error to the $p$'th power (with $p=2.5$) and the
normalized magnitude of the centered difference of the state versus
the time index.  The power of $p$ was chosen to aid in the
agreement between the two quantities.  It is clear that the two
quantities show a strong correlation.  Right: A log-log plot of the
centered time difference vs. the dynamic estimation error.}
\end{center}
\end{figure}

The left plot of Figure~\ref{fig:e_d_comp} provides a comparison
between the $p$'th power of the Kalman error (with $p=2.5$) and a
measure of the time variability in the MHD simulation, $d_m$, defined
by the centered difference: $d_m = \frac12\left\| \vec{x}_{m+1} -
\vec{x}_{m-1}\right\|$.  The Kalman error has been raised to the
$p$'th power to aid in the empirical agreement between the two
quantities.  Both $(e_m)^p$ and $d_m$ have been normalized by dividing
by their respective maximum for the purpose of displaying them
simultaneously in the left plot of Figure~\ref{fig:e_d_comp}.  The
right plot in Figure~\ref{fig:e_d_comp} shows $\log(d_m)$
vs. $\log(e_m)$ and the line of best fit.  From
Figure~\ref{fig:e_d_comp}, it is clear that $(e_m)^p$\ and $d_m$\ are
highly correlated, thus showing the relationship between temporal
variability and reconstruction error.

\section{Conclusion}

This paper has demonstrated promising results for the Kalman solution
of dynamic inverse imaging problems in which the object is evolving in
time as the data are being collected.  It was shown that the Kalman
solutions were superior to static reconstructions (based on sliding
time windows) except for time periods of extremely rapid variability,
in which case both methods were equally inadequate.  The Kalman
reconstructions presented here were implemented with two features to
improve the solutions:
\begin{enumerate}
\item{The solutions was regularized by replacing equation
(\ref{measurement}) with (\ref{measurement2}), which forces smooth solutions.}

\item{The random walk was given a correlation length with a
non-diagonal form of the state noise covariance matrix
$\mat{C}_{\vec{g}}$.}

\end{enumerate} 

There is much to be done along this
line of research.  Some of the significant issues that need to be
addressed are:
\begin{itemize}
\item{Enforcing positivity constraints: For many astrophysical
applications the unknown quantity to be imaged must be positive.  One
approach for this is suggested by Simon \& Chia (2002).}

\item{Reducing computational complexity: In order to solve large 2D
and 3D problems, the computational complexity must be reduced.  Much
work along these lines has already been done (e.g., Cane et al.\ 1996,
Asif \& Moura 1999, Treebushny \& Madsen 2003, Khellah et al.\ 2005,
and references therein) and it is an ongoing area of investigation.}

\item{Assimilating physics-based models: For solar work, using an MHD
model for the update operator in equation (\ref{time-update}) will
greatly increase the ability of the filter to track temporal changes.
For microquasar imaging in radio interferometry a kinematic model may
be suitable.}
\end{itemize}

The authors thank C.F. Gammie for the MHD simulation.  This research
was supported by NASA Sun-Earth Guest Investigator Program grant
NNG04GG32G to the University of Illinois.

\end{document}